\documentstyle[aps,graphics,amssymb,epsfig,eqsecnum,amsmath]{revtex}

\title{Comment on 'Invalidation of the Kelvin force in ferrofluids'}

\author{A. Engel\thanks{email:andreas.engel@physik.uni-magdeburg.de} 
          \\[2ex]
        {\small Institut f\"ur Theoretische Physik,
        Otto-von-Guericke Universit\"at, Postfach 4120, D-39016
        Magdeburg, Germany}}

\newcommand{\Ha}{{\bf H}}
\newcommand{\M}{{\bf M}}
\newcommand{\F}{{\bf F}}
\newcommand{\f}{{\bf f}}

\newcommand{\na}{\nabla}

\newcommand{\br}{\boldsymbol{r}}

\begin{document}

\maketitle

\vspace{0.5cm}


In a recent letter \cite{OdLi} Odenbach and Liu claim that their experimental
results for the force on a container filled with ferrofluid in an
inhomogeneous external magnetic field invalidate the standard 
Kelvin expression $\f={\mu_0 (\M\cdot\na)\Ha}$ for the magnetic force density
in a magnetizable medium. It is the purpose of this comment to point out that
the described experiment measuring the {\it total} force on a magnetizable
body cannot verify or falsify different expressions for the magnetic force
{\it density} without taking into account the corresponding surface
contributions. 

The magnetic force on a magnetizable body in thermodynamic equilibrium in an
external magnetic field can be determined in a simple and unambiguous way. The
change of the free energy of the body due to variations of the external field
is well-known to be \cite{LL}
\begin{equation}
  \delta F=-\mu_0\int_V d^3 r\; \M(\br)\cdot\delta\Ha_0(\br),
\end{equation}
where the integral is over the volume of the body, $\M(\br)$ is its local
magnetization, and $\Ha_0(\br)$ denotes the external field in the absence of
the body. If the change in the field is due to a displacement of the body by 
an infinitesimal vector $\delta \br$ we have $\delta\Ha_0=(\delta
\br\cdot\na)\;\Ha_0$. At the same time the corresponding change in free energy
is given by  $\delta F = -\F\cdot\delta \br$, where $\F$ is by definition the
total force on the body. Using $\na\times\Ha_0=0$ we find 
\begin{equation}\label{main}
  \F=\mu_0 \int_V d^3 r \;(\M\cdot\na)\;\Ha_0.
\end{equation}

This is a generally valid expression, subject only to the constraint of
thermodynamic equilibrium. In particular it does correctly describe the
experimental findings reported in \cite{OdLi}. 

By formal manipulations expression (\ref{main}) can be decomposed 
into a surface and a volume part in various ways. Besides the decomposition
advocated in \cite{OdLi} there is the standard possibility to use the Kelvin
force density $\f=\mu_0(\M\cdot\na)\;\Ha$ in the volume and a surface integral
over $\mu_0 M_n^2/2$ with $M_n$ denoting the normal component of the
magnetization \cite{Rosen}. If consistently used all these decompositions,
including the latter one using the Kelvin force density, are equivalent to
(\ref{main}) and therefore describe the experimental findings equally well.  

The expression for the force suggested in \cite{OdLi} (their eq.(6)) differs
from (\ref{main}) by a factor $(1+\chi)/(1+D\chi)$ with $D$ denoting the
demagnetization factor. This is probably due to the fact that at the same time
where demagnetization effects are taken into account also contributions from
the surface integral ( in their case involving $M_t^2$) matter. Since in
the experiment $D\cong0.9694$ the difference between their result (6) and the
correct expression (\ref{main}) is too small to cause noticable differences
with the experiment. 

In conclusion the main aim of the letter namely to invalidate the Kelvin force
on the basis of experimental facts was not accomplished. Moreover the variant
expression (6) offered as alternative to describe the experiment is incomplete
due to the neglect of surface contributions.\\

{\bf Acknowledgement:} Discussions with Hanns-Walter M\"uller and Ren\'e
Friedrichs are gratefully acknowledged.

\end{document}